\documentclass[10pt,preprint,eqsecnum]{aastex}
\usepackage{amsmath}

\begin{document}

\title{Supernova Enrichment of Dwarf Spheroidal Galaxies}

\author{P. Chris Fragile, Stephen D. Murray, Peter Anninos}
\affil{University of California,
Lawrence Livermore National Laboratory, P.O. Box 808, Livermore, CA 94550}

\and

\author{Douglas N. C. Lin}
\affil{University of California,
Lick Observatory, Santa Cruz, CA  95064}


\begin{abstract}
Many dwarf galaxies exhibit sub-Solar metallicities, with some star-to-star
variation, despite often containing multiple generations of stars.  
The total metal content in these systems is much less than
expected from the heavy element production of massive stars in
each episode of star formation.  Such a deficiency implies that a
substantial fraction of the enriched material has been lost from these
small galaxies.  Mass ejection from dwarf galaxies may have important
consequences for the evolution of the intergalactic medium and for the
evolution of massive galaxies, which themselves may have formed via
the merger of smaller systems.  We report here the results of
three-dimensional simulations of the evolution of supernova-enriched
gas within dwarf spheroidal galaxies (dSph's), with the aim of determining
the retention efficiency of supernova ejecta.  We consider two galaxy models,
selected to represent opposite ends of the dSph sequence.  One contains
$10^6$~M$_\odot$ of gas, and the other $5.5\times10^6$~M$_\odot$.  In both,
the baryonic-to-dark matter ratio is assumed to be 0.1.  
The total binding energies of the
gas in the two models are $9.8\times10^{50}$~erg and $1.6\times10^{52}$~erg.
For each model galaxy we investigate a number of scenarios.  The simplest is
a single supernova within a smooth gas distribution.   We also investigate 
the evolution of ten supernovae, within initially smooth gas distributions, 
occurring 
over time spans of either 10 or 100~Myr.  Finally, we investigate the effects
of ten supernovae occurring over 10~Myr in a medium filled with hot
``bubbles,'' such as would be expected in the presence of an initial
generation of hot stars.  For models with only a single supernova, no
enriched material is lost from the galaxies.  When multiple supernovae
occur within an initially smooth gas distributions, 
less than half of the enriched gas
is lost from the galaxy (fractional losses range from 0-47\%).  Most of the
enriched gas is lost, however, from the cores of the galaxies.  In the
presence of an initially disturbed gas distribution, 
6\% or less of the enriched gas
remains in the core, and much is lost from the galaxies as a whole (47\% and
71\% for the larger and smaller galaxy models, respectively).  If subsequent
star formation occurs predominantly within the core
where most of the residual gas is concentrated, then these results could
explain the poor self-enrichment efficiency observed in dwarf
galaxies.
\end{abstract}


\section{Introduction}
\label{sec:intro}

In cosmological models dominated by cold dark matter (CDM), the
amplitude of fluctuations increases toward shorter wavelengths.  Small
galaxies are therefore likely to be the first to form, and large
galaxies subsequently form through the merger of smaller systems
\citep[e.g.][]{WR78,BFPR84,C94,KNP97,NFW97}.  Small systems 
observed today, such as dwarf spheroidal
galaxies (dSph's), may thus represent surviving ``building blocks'' of
larger galaxies, and therefore much attention has been paid to
understanding the gas evolution and first generation of star formation
within them \citep[e.g.][]{LM94,NW94,M97,WEE98,K99,SLGV99,B00,MB00,
DML03}.

Their small sizes might indicate that dSph's can also serve as
relatively simple laboratories to study the effects of various
physical processes upon star formation.  In spite of their dynamical
simplicity, however, they are found to have a wide range of 
star-formation and enrichment histories, with few clear trends 
\citep{Mateo98,Grebel01,H01}.  The data therefore indicate
that multiple factors have affected star formation in dSph's, making
them important and interesting objects for the study of early star
formation in galaxies.  Factors that may have influenced star
formation in dSph's include their exposure to external ultraviolet radiation, 
the depth of their gravitational potential, their overall mass, their
interactions with other dwarf systems, and their proximity to neighboring
massive systems.  The work described here is part of a series of papers in
which we examine these various physical factors, with the goal of determining
their relative importance upon the evolution of the dwarf galaxies.

An important observational fact concerning dSph's is that
almost all have multiple generations of stars, with many
showing clear evidence of having undergone bursts of star formation,
generally with one dominant, and multiple weaker events \citep{Mateo98,
Grebel01}.  In some systems, multiple generations of star
formation may also be inferred from the observed spread in the
metallicity of individual stars within them \citep{Lehnert92,
Suntzeff93, Cote00, Shetrone01}.  Multiple generations of stars are
also inferred from Hubble Space Telescope observations of stellar
populations in dwarf systems that are resolved both spatially as well
as temporally \citep{Getal98, DPetal98a, DPetal98b, DPetal02}.
Most systems with multiple generations of stars, however, display less
self-enrichment than
would be expected from their star-formation histories, and the total
mass of heavy elements contained within these galaxies is far less than
that expected to be produced by the massive stars that would have formed
along with the currently observed low-mass stellar population 
\citep{S96,Grebel97}.

Star formation within dSph's may be strongly inhibited at early times
by the ultraviolet background radiation \citep{E92,QKE96,
KBS97,NS97,WHK97,BL99,BLBCF02}.  External triggers may still, however, lead to
cooling and star formation \citep{LM94,DML03}.  Following an initial burst
of star formation, the remaining gas within the galaxy may be heavily
affected by photionization, stellar winds, and supernova explosions.
Such stellar feedback has been studied extensively as a means of driving
the evolution of stellar systems.  Photoionization feedback has been
proposed as a means by which star formation may be self-regulated in dwarf
systems \citep{BR92, AB00, DML03}.  In systems the size of globular clusters,
photoionization-driven winds may lead to extensive mass loss, sufficient
even to completely terminate star formation \citep{PD68, TBLN86}.  In small
galactic systems as extended as dSph's, however, it is ineffective
at driving mass loss \citep{NBLT89}.

Energy input from the multiple supernova events following a burst of
star formation may also be a means
of driving mass loss.  Substantial mass loss has been found in simulations
of low-mass disk systems subjected to multiple supernovae (Mac Low \& Ferrara
1999).  Mass loss is expected to be less efficient, however, from more
rounded systems, in which heated gas must travel a greater distance
through the ambient medium before it is able to leave the system.  
Studies of the evolution of supernovae in high redshift dwarf systems
find that the high densities and resultant high cooling efficiencies
and external pressures make it relatively easy for dwarf galaxies to
retain mass following supernova explosions.  In order for them to
experience significant mass loss, high ($\gtrsim10$\%) star formation
efficiencies are required \citep{DS86,MLM88,MFR01,MFM02}.  Even at later
times, high star formation efficiencies may be required in order for 
supernovae to drive extensive mass loss, due to radiative losses
\citep{Larson74}.  We note, however, that mass loss through the
Lagrange points may also lead to significant rates of mass loss, if the
gas within the galaxy is ionized either by external or internal sources
\citep{MDL03}.  Extensive star formation is, therefore, not required in 
order to account for the total removal of gas from dSph's, as observed today.

The presence of multiple generations of stars within dSph's implies
that their early star formation efficiencies were not so high as to
lead to extensive mass loss.  The poor self-enrichment of later generations, 
however, suggests that enriched material from
supernovae was preferentially lost, leaving behind gas that was at most
slightly enriched.  It was from this poorly enriched gas that 
subsequent generations of stars then
formed.  It is therefore of interest to examine the
evolution of enriched material within dwarf galaxies that have
undergone relatively inefficient bursts of star formation.
In this paper, we examine the loss of enriched material from
dSph's which are subjected to a number of supernova events.  We
proceed in \S~\ref{sec:method} by describing the numerical method and
the setup of the models.  The results of the simulations are described
in \S~\ref{sec:results} and \S~\ref{sec:differences}.  The possible
consequences for the evolution of dwarf galaxies are discussed in
\S~\ref{sec:consequences}.

\section{Numerical Method}
\label{sec:method}

\subsection{The Numerical Code}
\label{sec:methodcode}

The models discussed below have been computed using Cosmos, a
massively parallel,
multidimensional, radiation-chemo-hydrodynamics code 
for both Newtonian and relativistic flows developed at Lawrence
Livermore National Laboratory.  The relativistic capabilities and
tests of Cosmos are discussed in \citet{AF03}.  Tests of the Newtonian
hydrodynamics options and of the internal physics relevant to the
current work are presented in \citet{AFM02}, 
and so we shall not discuss those in detail here.

Because we shall be 
examining the effects of having multiple supernova explosions at
random locations throughout the cores of dwarf galaxies, we cannot
make any assumptions regarding the
symmetry of the systems. As a consequence, our models are run in
three-dimensions on a Cartesian mesh.  Most of the simulations 
are run with resolutions of
$128^3$ zones.

\subsection{Radiative Cooling}
\label{sec:methodcooling}

Chemistry is not followed in these simulations.  Instead, local
cooling is given by the following cooling function, based in part on an
equilibrium (hydrogen recombination and collisional excitation)
cooling curve
\begin{equation}
\Lambda(T,\rho) = \left[\sum_i \dot{e}_i(T) (f_I \rho)^2 + \dot{e}_{M}(T)
(f_M \rho)^2 \right]
    \times \left\{ \begin{array}{ll}
                   e^{(T-T_{min})/\delta T} & \mbox{if} ~ T \le T_{min} , \\
                   1 & \mbox{otherwise} ,
                   \end{array}
           \right.
\label{eqn:coolingnochem}
\end{equation}
where $\dot{e}_M$ is the temperature-dependent cooling rate from
metals (including carbon, oxygen, neon, and iron lines, assuming a
metallicity $Z=0.03Z_\odot$), $\dot{e}_i$ is the cooling rate from
hydrogen and helium lines, $f_M$ is the mass fraction of metals taken
to be $6 \times 10^{-4}$, $f_I$ is an estimate of the ionization
fraction, defined as $\mbox{min}(1,~\mbox{max}(0,~(T_{eV} - T_c)/3) )$
with $T_c=1 eV$ to match roughly the upper and lower bounds in the
equilibrium model described in \citet{AFM02}, $T = (\Gamma-1)
e\mu m_p/(k_B \rho (1+f_I))$ is the gas temperature in
Kelvin, $\rho$ is the gas density, $e$ is the internal energy density
of the gas, $\mu$ is the mean molecular weight, taken to be unity, $m_p$ is
the proton mass, and $k_B$ is Boltzmann's constant.  The exponential
is introduced to suppress cooling at low temperature (here
$T_{min}=10^4$ K, with width $\delta T = 500$ K). The initial
temperature of the gas in the systems considered here is 8200~K,
representing a state which is partially ionized by the background.
The rapid decrease of cooling below $10^4$~K simulates the effects of
heat sources and the exponential decay of electron density
below about 1 eV in equilibrium models.
This prevents the gas from radiatively cooling significantly below its
initial temperature.

Our choice of metallicity is motivated by the fact that we are
examining the evolution of young dSph's, which have not had a chance
to undergo significant self-enrichment.  The value $Z=0.03Z_\odot$, is
appropriate to the older generations of stars observed within
dSph's \citep{Grebel01}.

\subsection{Gravity}
\label{sec:methodgravity}

The galaxies are modeled as non-self gravitating gas within a fixed 
dark-matter potential.  The omission of self-gravity is reasonable,
given that the
gravitational potential of dwarf galaxies with $M_g\lesssim10^9 M_\odot$
is dominated by the dark-matter halo \citep{PSS96}, where $M_g$ is 
the mass of the 
visible (gas plus stellar) matter.  The shape of the potential is 
given by a modified Hubble profile (Binney \& Tremaine 1987)
\begin{equation}
\phi(r<R_t) = \frac{G \tilde{M}}{R_c} \left\{ 1 - \frac{\mbox{ln}
[x + (1+x^2)^{1/2}]}{x} \right\} ~,
\end{equation}
where $x=r/R_c$, $R_c$ is the core radius, and
\begin{equation}
\tilde{M} = M_d \left\{ \mbox{ln}[x_t +(1+x_t^2)^{1/2}] - 
x_t(1+x_t^2)^{-1/2} \right\}^{-1} ~,
\end{equation}
where $M_d$ is the dark-matter mass, $x_t=R_t/R_c$, and $R_t$ is the 
tidal radius.  This potential has a simple form,
while retaining the essential behavior for the shape of potentials which are
generally observed in dSph's (Burkert 1995).  The gas is initialized to be
isothermal and in hydrostatic equilibrium within the potential, such that the
density
\begin{equation}
\rho =\rho_0 e^{-\phi/c_s^2} ~,
\label{eqn:densitylaw}
\end{equation}
where $c_s$ is the isothermal sound speed.  The value of $\rho_0$ is scaled to
give the desired gas mass $M_g$ within $R_t$.  The effects of a tidal
radius are included by flattening the potential beyond $R_t$.  
Beyond $R_t$, the gas density is also decreased by a factor of
0.01, and the temperature is correspondingly increased in order to maintain
hydrostatic equilibrium.  This simplified treatment of the galaxy's
surroundings allows us to focus on the physics within the dwarf galaxy
itself.

\subsection{Boundary Conditions}
\label{sec:methodboundary}

To minimize the effects of the computational boundaries on the 
galaxy evolution, we choose the size of the grid, $l$, such that 
$2R_t = 0.8l$.  Our boundary zones remain static throughout these
simulations, maintaining their initial density and temperature.
This choice of boundary conditions resulted in fewer disruptive
reflections than flat (vanishing first derivative) boundaries
and less spurious mass loss than outflow boundaries.  
We have verified that, in the absence of
supernovae, the models remain static for several dynamical times.

\subsection{Galaxy Parameters}
\label{sec:methodgalaxy}

\begin{figure}
\plotone{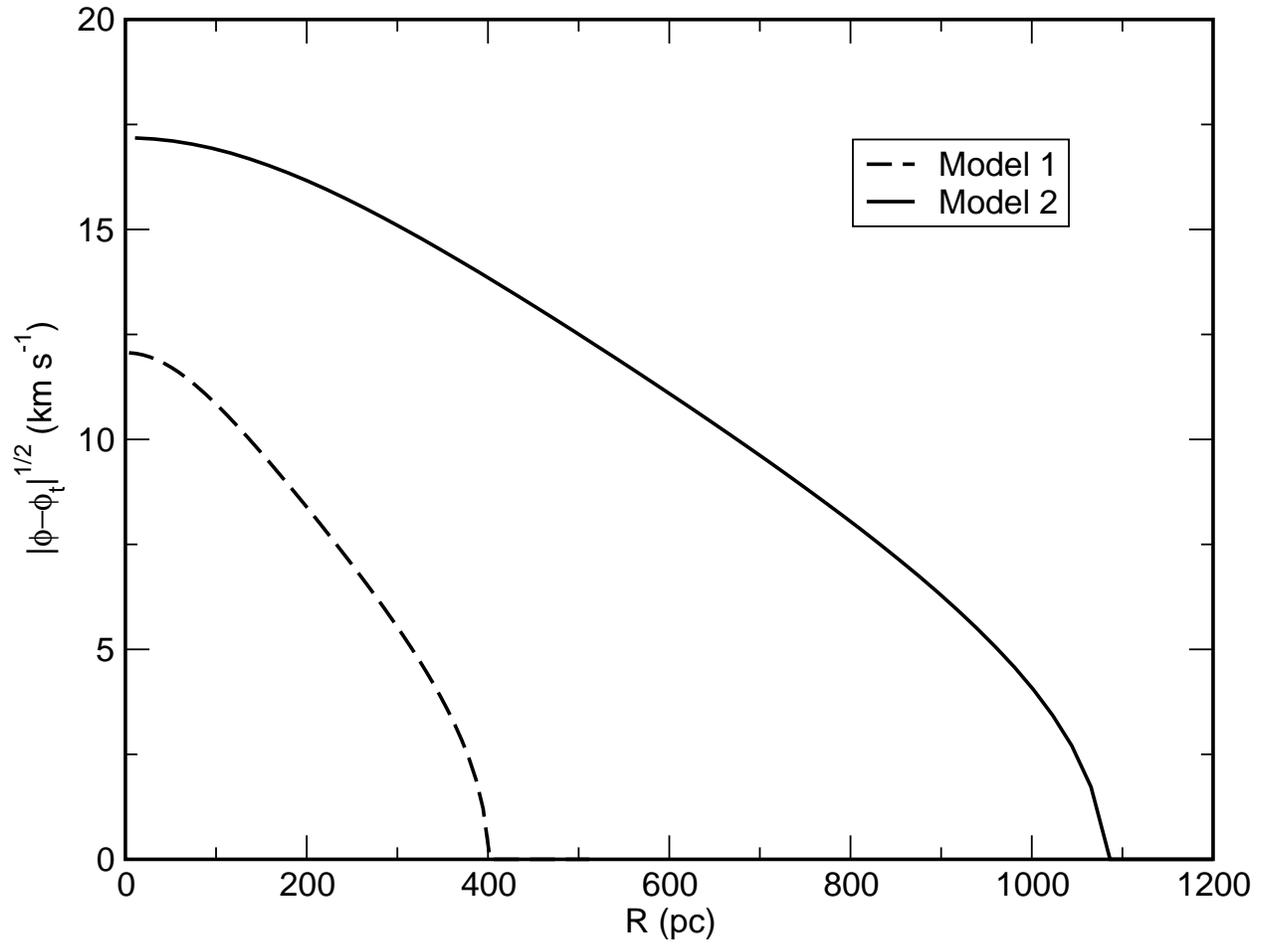}
\caption{Square root of the absolute value of the normalized gravitational
potential as a function of radius for galaxy model 1 (dashed curve)
and galaxy model 2 (solid curve).
}
\label{fig:potential}
\end{figure}

Two different model galaxies have been examined.  Their parameters are given
in Table~\ref{tab:galaxies}, which lists the dark-matter mass $M_d$, gas mass
$M_g$, core radius $R_c$, the central gas density $\rho_0$, the dynamical time
of the galaxy $\tau_d$ (defined below), the total gravitational binding energy
of the gas, $\Phi$, and the total thermal energy of the gas, $E_{th}$,
for each model.  The parameters of galaxies 1 and
2 are chosen to be close to those of Leo~II and Fornax, respectively,
objects near the ends of the distribution for dSph's (Mateo 1998).  For both
model galaxies, the ratio of the tidal to the core radii 
is chosen to be $R_t/R_c=3$.  We note that
the central densities of the two models are similar, in spite of their
having different masses, due to the larger core radius of Model~2. In
Figure~\ref{fig:potential}, we plot the square root of the absolute 
value of the normalized gravitational potential 
($\vert \phi - \phi_t \vert^{1/2}$) as a function of radius 
for the two galaxy models.  This function gives a good estimate of the
local escape velocity.

It may also be noted from Table~\ref{tab:galaxies} that the total 
binding energy of the gas in the two models is of the order of the
energy input of 1-10 supernovae, which is precisely the range covered 
in our models (see below).  It might therefore be expected that the
energy input from supernovae would lead to extensive mass loss from the
systems.  As shall be seen, however, this is only the case when radiative 
cooling is neglected.  As has been noted in earlier
studies \citep{Larson74}, radiative losses cause 
the deposition of energy from
supernovae to be relatively inefficient.  We discuss this point further 
and examine the effects of
radiative losses in \S~\ref{sec:10SNnobub}.

\subsection{Treatment of Supernovae}
\label{sec:methodsupernovae}

In this study, 
supernovae are simulated by adding 10$^{51}$ erg of internal energy
to the gas over finite regions.  The internal energy is injected
over an approximately spherical volume with a ``top-hat'' profile
of 3-4 zones radius.  Computing time constraints limit our
resolutions, and we are therefore unable to maintain zoning sufficiently
fine as to resolve the supernova shocks and the cooling regions
immediately behind them.  The results are therefore vulnerable to
numerical over-cooling of the shocked gas, which would lead to an
underestimate of the effects of the supernovae.  This potential
problem is exacerbated by having the energy initially deposited into
the internal energy of the gas, after which it evolves to a
Sedov-Taylor solution.  
In addition, the artificial viscosity used to capture shocks 
spreads out shock heating
over a few zones.  Gas may, therefore, spend multiple timesteps at elevated
temperatures, possibly near the peak of the cooling curve.  This can
cause further over-cooling of the gas, significantly affecting the 
hydrodynamic evolution.  In order to prevent over-cooling of the gas we 
simultaneously employ two techniques.  First, we do
not allow cooling within a supernova until is has roughly doubled in
radius.  Second, we follow the common procedure of 
preventing cooling of the gas during shock passage \citep{SSM85, BFK90, AN94}.
To accomplish this in a three-dimensional problem, 
where shocks may occur in any direction, a
simple scalar criterion is needed.  We find, in practice, that the ratio 
of the artificial viscous heating term to gas pressure is an adequate 
filter.  Thus, cooling is turned off whenever $Q/P > 0.1$.  The effects 
of these techniques will be explored in \S~\ref{sec:10SNnobub}.

In order to follow the evolution of the supernova-enriched material, we add a 
tracer fluid to the zones in which the supernova energy
is initially deposited.  The tracer is a
passive quantity that is advected in the same manner as density.

The simulations discussed below include either one or ten supernovae.  In the
cases with only a single supernova, it is set off either at the center of the
system or at the core radius.  In these cases we are able to take
advantage of symmetries in the problem to reduce the required computational
resources.  For a single supernova set off at the center of the galaxy, we
employ $x$, $y$, and $z$ symmetries and run a single octant with reflective
boundary conditions at the $x=0$, $y=0$, and $z=0$ planes.  
For a single supernova at
$R_c$, we can invoke symmetry in two dimensions and run a quadrant with
reflective boundaries along two planes.  
If ten supernovae are included, the first is set off at $R_c$ at the start
of the simulation.  The rest are chosen to occur at
random orientations to the
center of the system.  The radii at which they occur are chosen at random,
weighted by the enclosed mass of gas, and constrained to be within $R_c$.
The latter constraint is due to the expectation that bursts of star 
formation are expected to be located predominantly in 
galactic cores, where gas 
densities are highest \citep{DML03}.  The timescale over which
supernovae occur is
determined by the timescale of the burst of star formation that led to the
formation of the progenitor stars, and by the spread in masses, and therefore
lifetimes, of the progenitor stars.  To examine the effects of reasonable
variations in these factors, the supernovae in the models are set off at
times chosen randomly within total spans of either 10 or 100 Myr.  

\subsection{The Form of the Interstellar Medium}
\label{sec:methodism}

In most of our models, the gas density varies smoothly, following
Equation~\ref{eqn:densitylaw}.  Following a burst of star formation in
a dwarf galaxy, however, such a smooth density distribution is
unlikely to actually be found.  Instead, ionizing photons and winds
from massive stars, and turbulent motions shall all lead to a very
irregular gas distribution for the interstellar medium (ISM).  The low
density regions in such a distribution may enhance the loss of
material from supernovae, and so it might be expected that the form of
the gas distribution could play a role in the ability of a galaxy to
retain enriched gas.  In order to explore the role of the gas
distribution, two of the models include 100 ``bubbles,'' each five
zones in radius, whose density is decreased by a factor of five
relative to the surrounding gas.  The physical interpretation of these
bubbles is that they are cleared by ionizing photons and winds from
massive stars.  The temperature within the bubbles is increased, such
that they remain in pressure equilibrium with the surrounding gas.  
The bubbles are selected to
have random orientations around the center of the galaxy and they are
not, in general, co-located with the supernovae, which is justified 
by the fact that during their main-sequence 
life span, massive stars may move up to 100 pc from their
initial position.  The distances of the bubbles from the galaxy center
are selected at random, weighted by enclosed mass and constrained to
be no more than 0.6$R_t$ from the center of the galaxy.  Once the bubbles are
placed, the density of
the entire system is increased by a common factor to preserve the
original mass of the galaxy.  To simulate the presence and evolution of the
hot stars causing the bubbles, cooling is not allowed within the bubbles for
0.5 dynamical times after the supernovae end.

In physical units, the bubble radii correspond to approximately 40~pc for
Model~1 and 100~pc for
Model~2, comparable to the radii of Str\"omgren spheres.
The total filling factor for the bubbles within $0.6R_t$ is approximately
30\%, i.e. the ISM is highly irregular in form.
For typical stellar initial mass functions, the number of supernovae and
bubbles in our models represent overall star formation efficiencies for the
galaxies of less than 1\%.

\subsection{Other Physics}

The models discussed below do not include the effects of conduction or of
magnetic fields.  The effects of thermal conduction may be estimated from
the conductive flux
\begin{equation}
F_c=\kappa T^{5/2} \nabla T,
\end{equation}
where $\kappa=6\times10^{-7}$ erg cm$^{-2}$ s$^{-1}$ K$^{-7/2}$ 
\citep{Spitzer56}.  The total energy
flow across the interface between an expanding bubble of heated gas may be
compared with the radiative energy losses from within the volume of the
bubble.  Radiative losses rapidly lead to a relatively shallow temperature
gradient, greatly reducing the conductive flux, with the result that
conductive losses are found to be much less important than radiative emission
in our models, justifying the omission of conduction.

The presence of even weak magnetic fields would greatly hamper conduction,
further reducing its importance \citep{RT89}.  Stronger magnetic fields
could potentially affect the hydrodynamics.  A strong, ordered magnetic
field would tend to direct the flow of gas to be one-dimensional, whereas a
strong field that was tangled on small scales would act as an additional 
pressure term, possibly helping to contain gas within the galaxies.  The
magnetic field would be expected to noticeably affect the flow if the
magnetic pressure were comparable to the ambient gas pressure, which, for
our models, requires $B\approx5\mu$G, greater than or comparable to values
measured in the local ISM \citep{Heiles87}.  Whether dwarf systems could
generate fields with strengths comparable to those in our own Galaxy is
uncertain, as are the actual field strengths present today in dwarf galaxies.
Magnetic fields therefore remain an uncertain quantity.  If present in 
young dSph's at magnitudes comparable to the local ISM, then their affect
would be expected to be comparable to that of the ambient gas pressure,
having a quantitative affect upon our results, but not affecting the 
qualitative conclusion that enriched material can be preferentially 
lost from dwarf systems.

\section{Model Results}
\label{sec:results}

The parameters of the models that we have computed are shown in
Table~\ref{tab:models}.  Listed in the table for each model are the galaxy
model used, the number of supernovae included $N_{SN}$, the timescale over
which the supernovae occur $\tau_{SN}$, and the number of bubbles included
in the galaxy $N_{\rm{bub}}$.  Variations of many of these base models 
are also examined (see 
below).  All of the runs are summarized in Table \ref{tab:results}, where the 
variations are indicated by numbers following the letters of the base 
models.  The tabulated results include 
the percentage of tracer lost from the core as well as that lost beyond
$R_t$, and also the percentage of gas mass lost from the galaxies.  The
core tracer loss is measured after at least 2.5 dynamical times for the
galaxies, while the tracer and mass losses beyond $R_t$ are estimated from
the global minima (see below).  In this section,
we discuss the results of the models, and we follow with a discussion of
the reasons for the differences among them in \S~\ref{sec:differences}.

\subsection{Single Supernova Events}
\label{sec:1SN}

In Models~A1 and B1, single supernovae are set off in the centers of
the two model galaxies.  For both models, all of the tracer
remains within the core of the galaxy.  The lack of tracer loss is expected,
given the core masses of the low and high mass models, which are approximately 
$1.6\times10^5$ and $3\times10^6$~M$_\odot$, respectively.
Such large masses of gas have no trouble containing the ejecta from
single supernova events.

We have also run versions of Models~A and B, in which a single supernova
was triggered at $R_c$ (Models A2 and B2, respectively).  
Although this obviously resulted in significant loss
of tracer from the core, only the low mass galaxy (Model~1) showed any loss of
tracer from the system (30\%).  
We conclude from this that multiple supernovae appear to be 
required in order for large fractions of enriched material to be lost
from within rounded galaxies, even as small as those considered here.

\subsection{Multiple Supernovae in Undisturbed Systems}
\label{sec:10SNnobub}

\subsubsection{Low Mass Galaxy Models}

In Models C and D, 10 supernovae are set off at random 
locations within the core
of the initially undisturbed galaxy Model~1.  The supernovae are triggered at
random times over a span of 10~Myr for Model~C and 100~Myr for Model~D.

We further utilize Model C to test the effects of 
resolution upon our results and 
to explore the impact of different choices for radiative cooling 
and metallicity.  Model C1 is our control model, as it 
utilizes our standard cooling options, resolution, and metallicity.  
Hence, this model is most comparable to Models A, B, D, E, F, G, and H.
Model C2 utilizes 
the same cooling options, but has a spatial resolution reduced by a factor 
of two.  Model C3 tests the effect of not shutting off cooling 
at shock fronts by ignoring the $Q/P>0.1$ criterion described 
in \S~\ref{sec:methodsupernovae}.  Similarly, Model C4 
tests the effect of not shutting off cooling 
in the initial supernova region by ignoring that requirement.  Model C5 
ignores both the $Q/P$ condition and the requirement to shut off cooling 
in the initial supernova region, and thus represents results with 
cooling turned on ``everywhere,'' and so likely suffers from significant 
over-cooling.  Model~C6 tests the effect of our 
choice of metallicity, by using a metallicity $Z=0.3Z_\odot$, 
an order of magnitude higher than that of Model~C1.
Finally Model C7 represents the extreme case in which
radiative cooling is turned off everywhere, and the gas may cool only
by adiabatic expansion.  We proceed by first
comparing the results of Models~C1 and D.  The effects of the variations
of Model~C are discussed in \S~\ref{sec:numericaleffects}.

\begin{figure}
\plotone{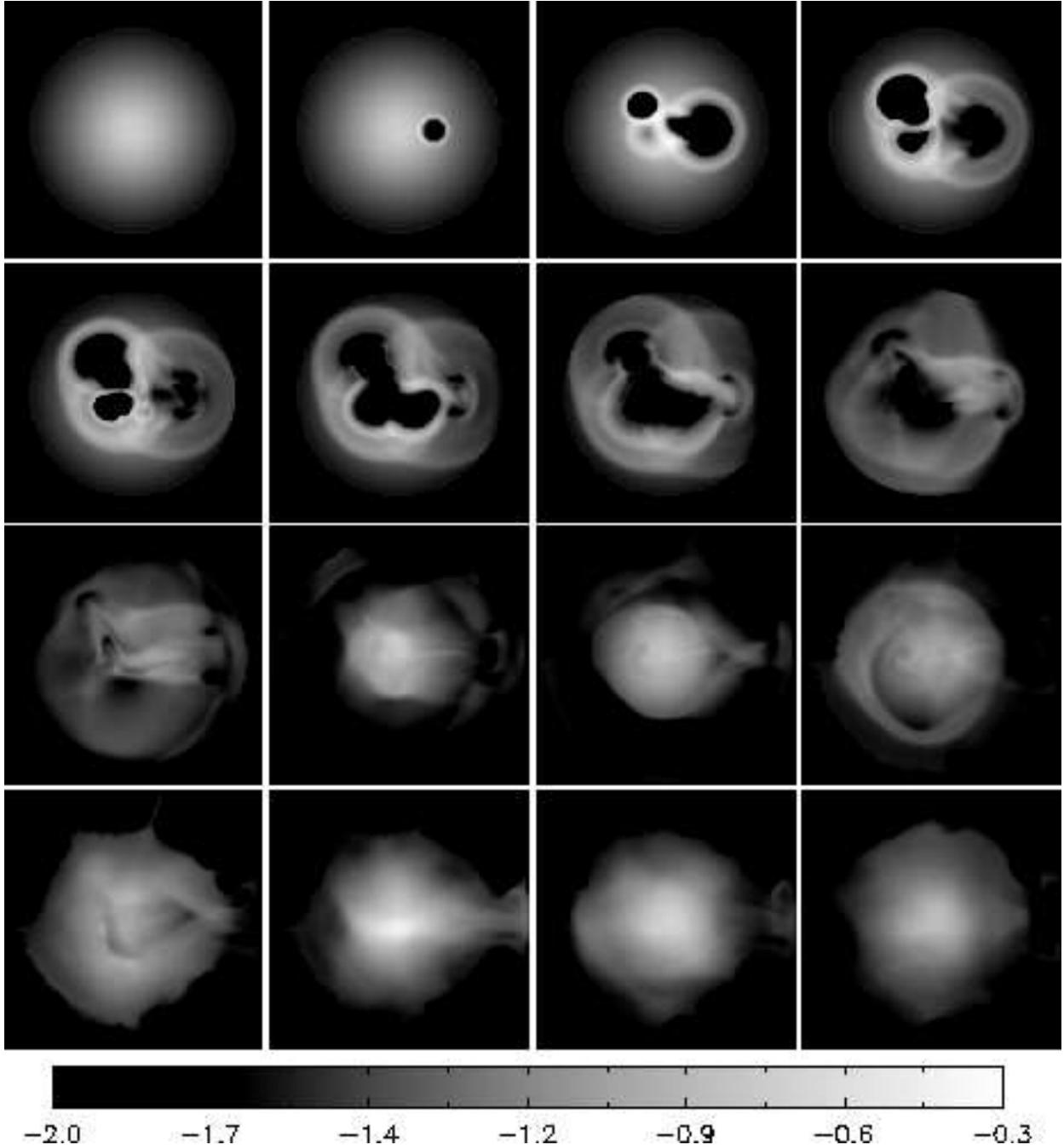}
\caption{Slices through the central plane of the galaxy showing the 
evolution of the logarithm of gas density for Model~C1 at times
0, 0.24, 3.8, 7.7, 9.6, 12.1, 14.5, 19.3, 24.1, 33.7, 43.4, 53.0,
62.7, 72.3, 81.9, and 96.4~Myr.  The units for the density scale in
this plot are $5.9 \times 10^{-24}$ g cm$^{-3}$.
}
\label{fig:rho10SNLeo10Myr}
\end{figure}

\begin{figure}
\plotone{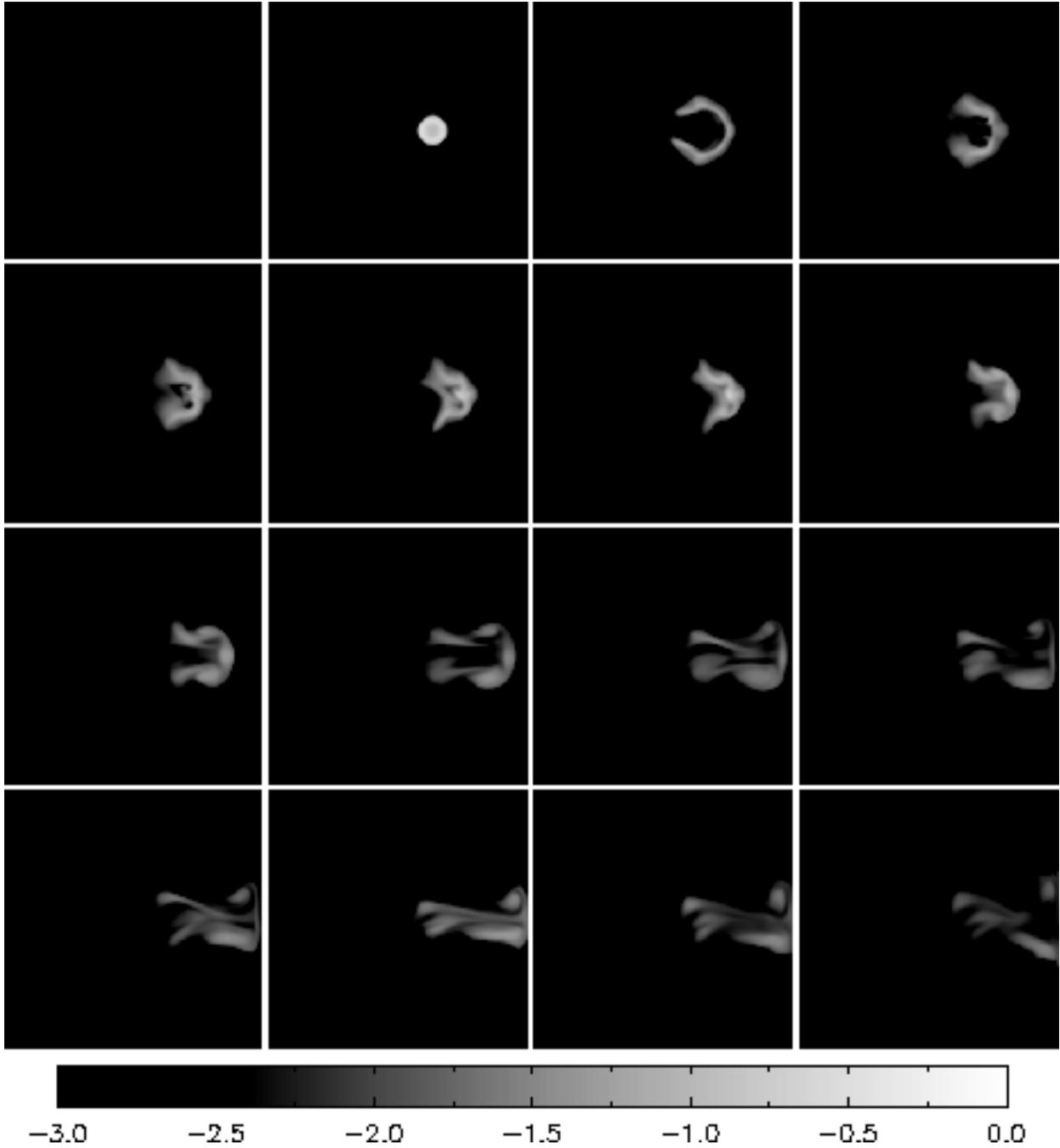}
\caption{Slices through the central plane of the galaxy showing the 
evolution of a single tracer fluid for Model~C1.  The frames correspond
to those shown in Figure \ref{fig:rho10SNLeo10Myr}.  The tracer fluid
belongs to the first supernova (triggered at $R_c$), visible in the 
second frame of 
Figure \ref{fig:rho10SNLeo10Myr}.  The logarithmic scale shows the 
relative amount
of tracer found in different regions.
}
\label{fig:tr10SNLeo10Myr}
\end{figure}

\begin{figure}
\plottwo{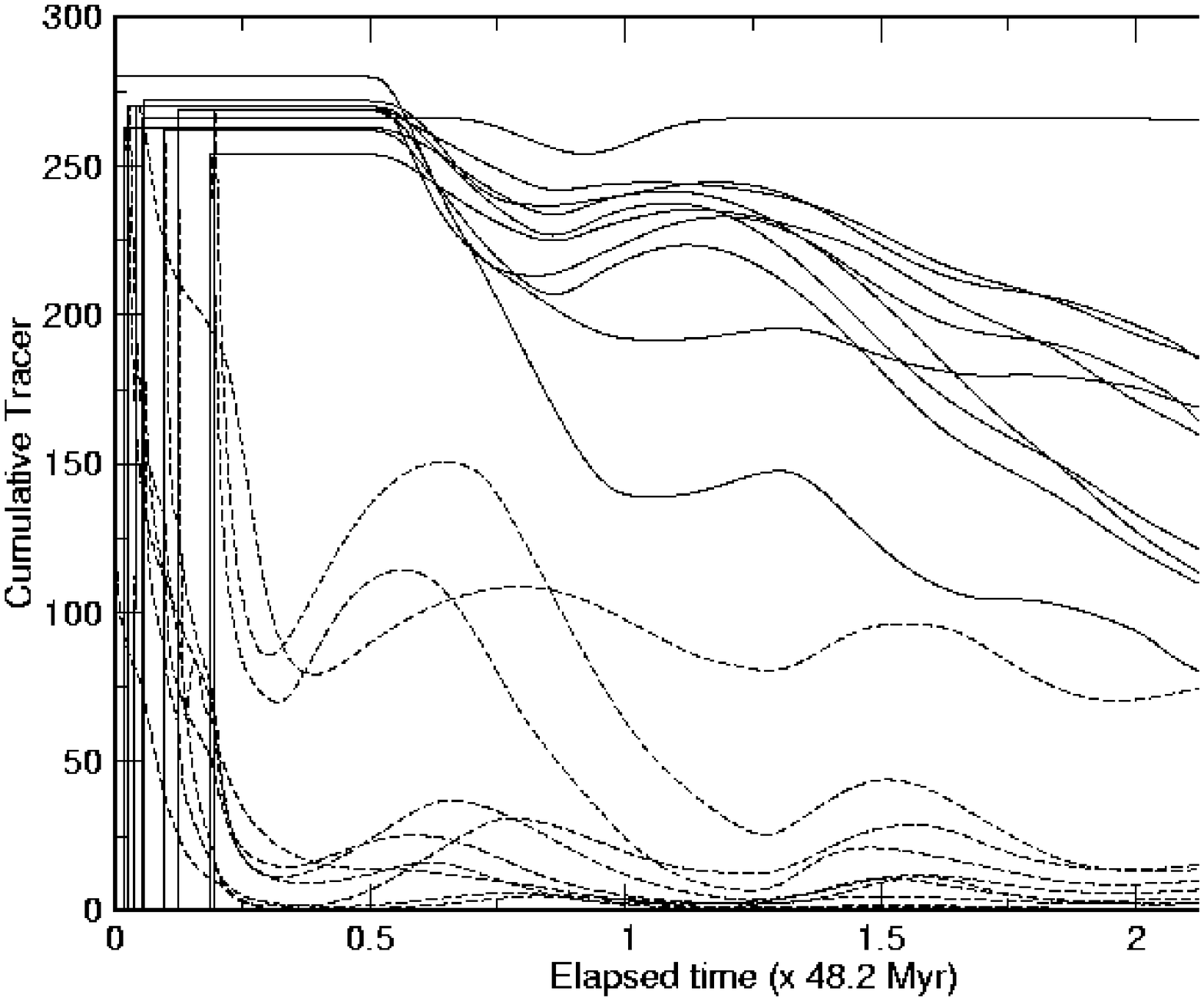}{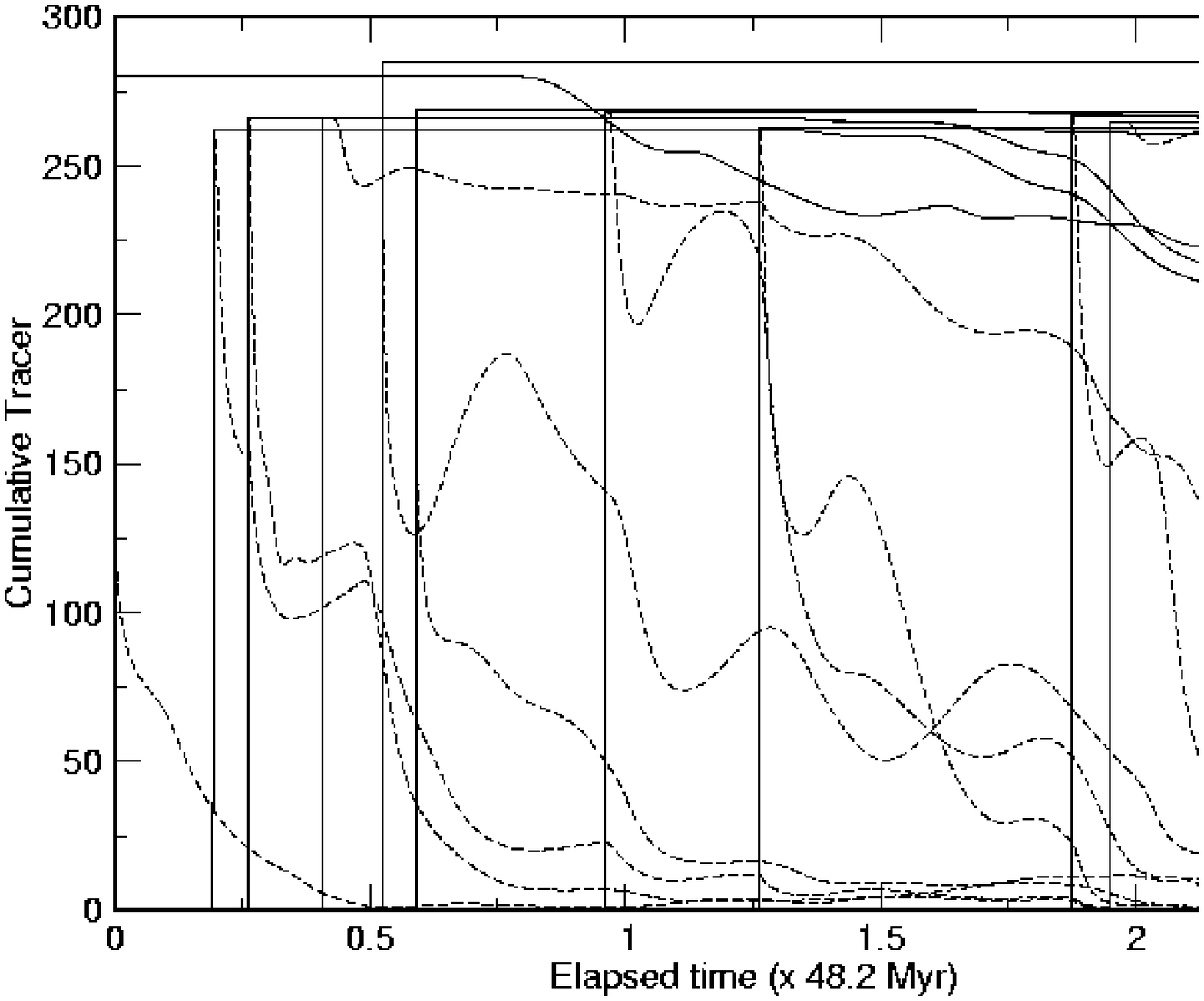}
\caption{Tracer evolution for (a) Model~C1 and (b) Model~D.
Shown are the amount of tracer for each supernova contained within the
tidal radius (solid curves) and within the core radius (dashed curves)
as a function of the elapsed time. 
}
\label{fig:trrad10SNLeo}
\end{figure}

In Figure~\ref{fig:rho10SNLeo10Myr} are shown snapshots of two-dimensional
slices of the density evolution through the central galactic plane for 
Model~C1.  The snapshots cover a span of approximately 100 Myr.  
The evolution of one of the tracer fluids from
Model C1 is shown in Figure~\ref{fig:tr10SNLeo10Myr}, at times corresponding
to the frames in Figure~\ref{fig:rho10SNLeo10Myr}.  This tracer is associated
with the initial supernova (triggered at $R_c$), visible in the second 
frame of Figure
\ref{fig:rho10SNLeo10Myr}.  
Figure~\ref{fig:trrad10SNLeo} shows the time evolution of the amount of tracer
contained within $R_t$ (solid curves) and $R_c$ (dashed curves) for Models~C1
and D.
The enriched material from each supernova is labeled with a different tracer,
and the evolution is shown individually for each.  On the
curves, time is presented in code units, where a unit of one represents 
approximately the dynamical time of the galaxy, defined as
\begin{equation}
\tau_d\equiv \frac{R_t}{c_s},
\label{eqn:taudyn}
\end{equation}
where $c_s$ is the sound speed of the gas.  The time unit for galaxy
Model 1 is 48.2 Myr.

Significant ($\gtrsim 80$\%) loss of tracer is seen from within $R_c$ for
both models, although  Model~C1 loses a greater fraction than 
does Model~D.  In Model~C1, the loss of tracer is accompanied by an initial
loss of approximately 50\% of the gas from the core.  The core mass returns
to its initial value, however, within $\approx\tau_d$.  This recovery 
of the initial core mass without an accompanying recovery of tracer 
implies that the tracer
does not mix well with the gas in the core.  In Model~D, where the supernovae
occur over a longer timescale, the evolution of the core mass is much
gentler, yet the same conclusion holds.

Additionally, in Model~C1, nearly one-half (47\%) of the tracer moves beyond
$R_t$, while almost 20\% does so in Model~D.  This material also does not mix
well with the surrounding gas, as shown in Figure~\ref{fig:tr10SNLeo10Myr},
where the tracer can be seen to evolve as a nearly cohesive unit.
As can be seen in Figure~\ref{fig:trrad10SNLeo} (and subsequent figures), 
some of the tracer that
moves beyond $R_t$ later returns to the galaxy.  This behavior is an artifact
of our boundary conditions.  At the boundaries,
the gas properties are held constant in time, which can artificially prevent
material from leaving the grid, and drive it back to the galaxy.
In actuality, material that moves beyond the tidal radius is unlikely to return
to the dwarf galaxy on the time scales shown here, but shall instead 
follow a separate orbit around the
parent galaxy.  The gas which escapes from the galaxy in Model~C1 does so with
velocities of only 5-7 km~s$^{-1}$, much less than the orbital speed of the
dSph satellites around the Milky Way.  During an orbit, therefore, 
the gas would
not move substantially away from the dSph, and would possibly 
re-encounter it on the next apogalacticon passage (Dong, Lin, \& Oh 2002).

\begin{figure}
\plottwo{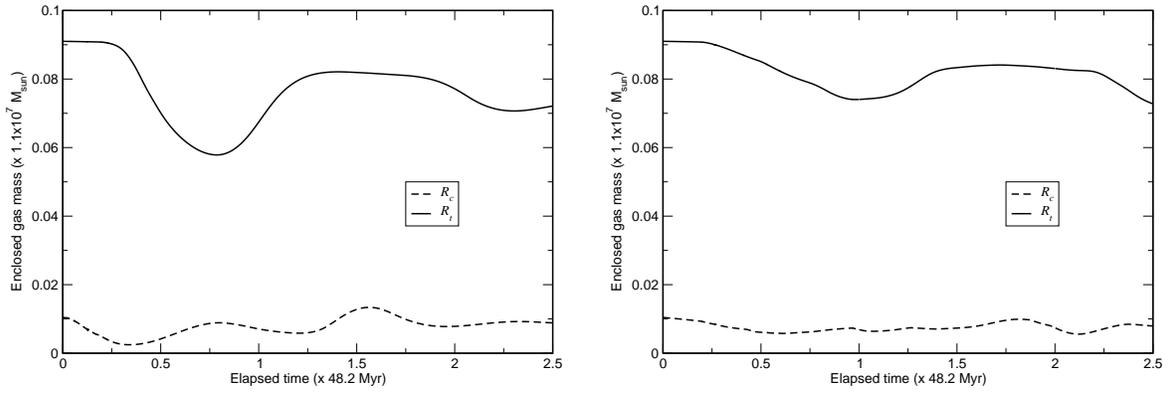}{f5b.eps}
\caption{Mass evolution of (a) Model~C1 and (b) Model~D.  Shown for each
model is the gas mass within the tidal radius $R_t$ (solid curves)
and the core radius $R_c$ (dashed 
curves) as a function of time.
}
\label{fig:mass10SNLeo}
\end{figure}

The loss of tracer from the galaxy is accompanied by overall mass loss, as
can be seen in
Figure~\ref{fig:mass10SNLeo}, which shows the mass contained within $R_t$
(solid curves) and $R_c$ (dashed curves) for Models~C1 and D.  The inflow due
to the outer boundary conditions is apparent in both figures.  As discussed
above, in the absence of a confining external medium, gas is actually unlikely
to return to the system after it moves to
beyond $R_t$.  The total mass loss to be expected in the models may therefore
be represented by the first minima shown in Figure~\ref{fig:mass10SNLeo}, which
represent 36\% and 21\% for Models~C1 and D, respectively.

\subsubsection{High Mass Galaxy Models}

\begin{figure}
\plottwo{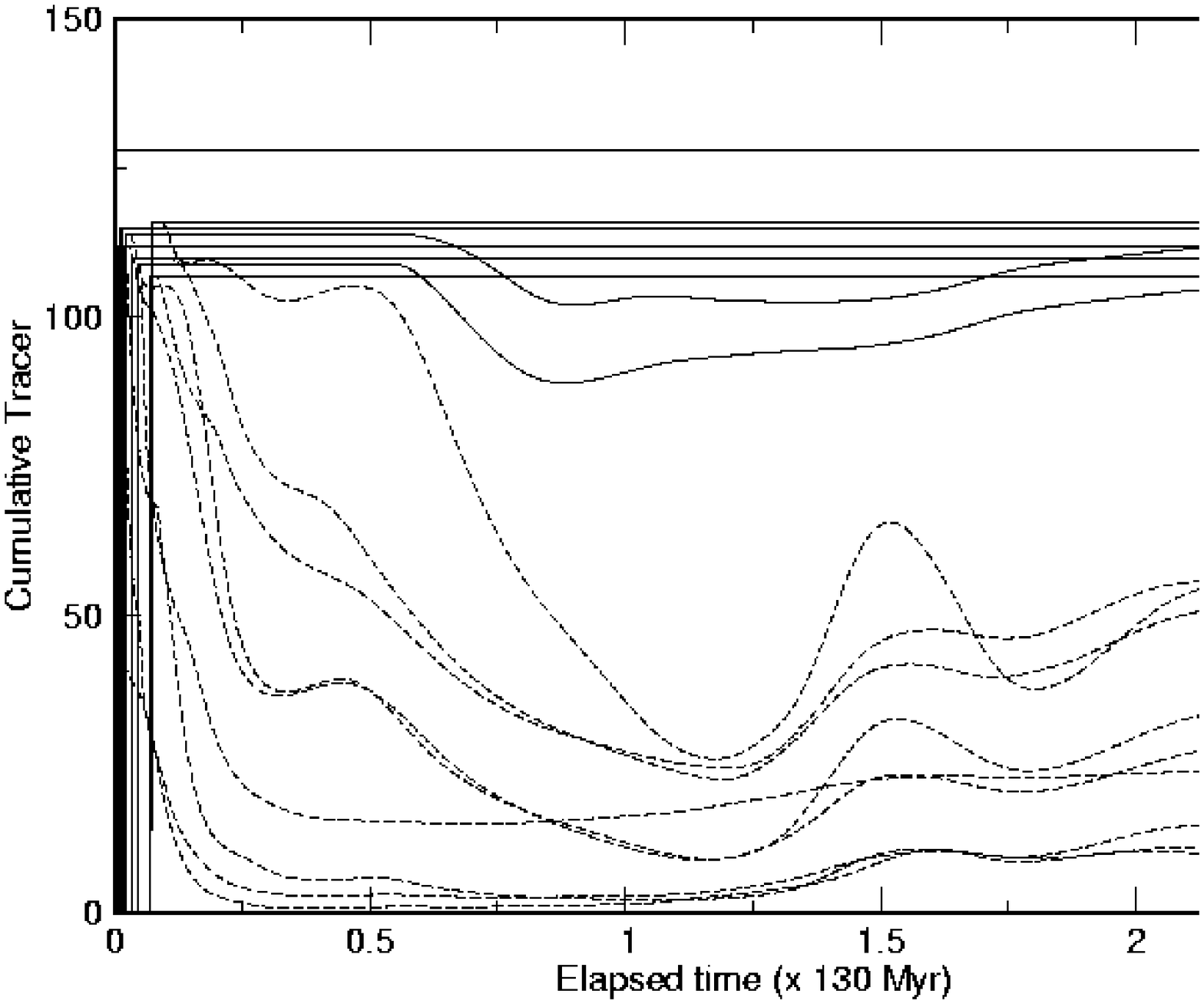}{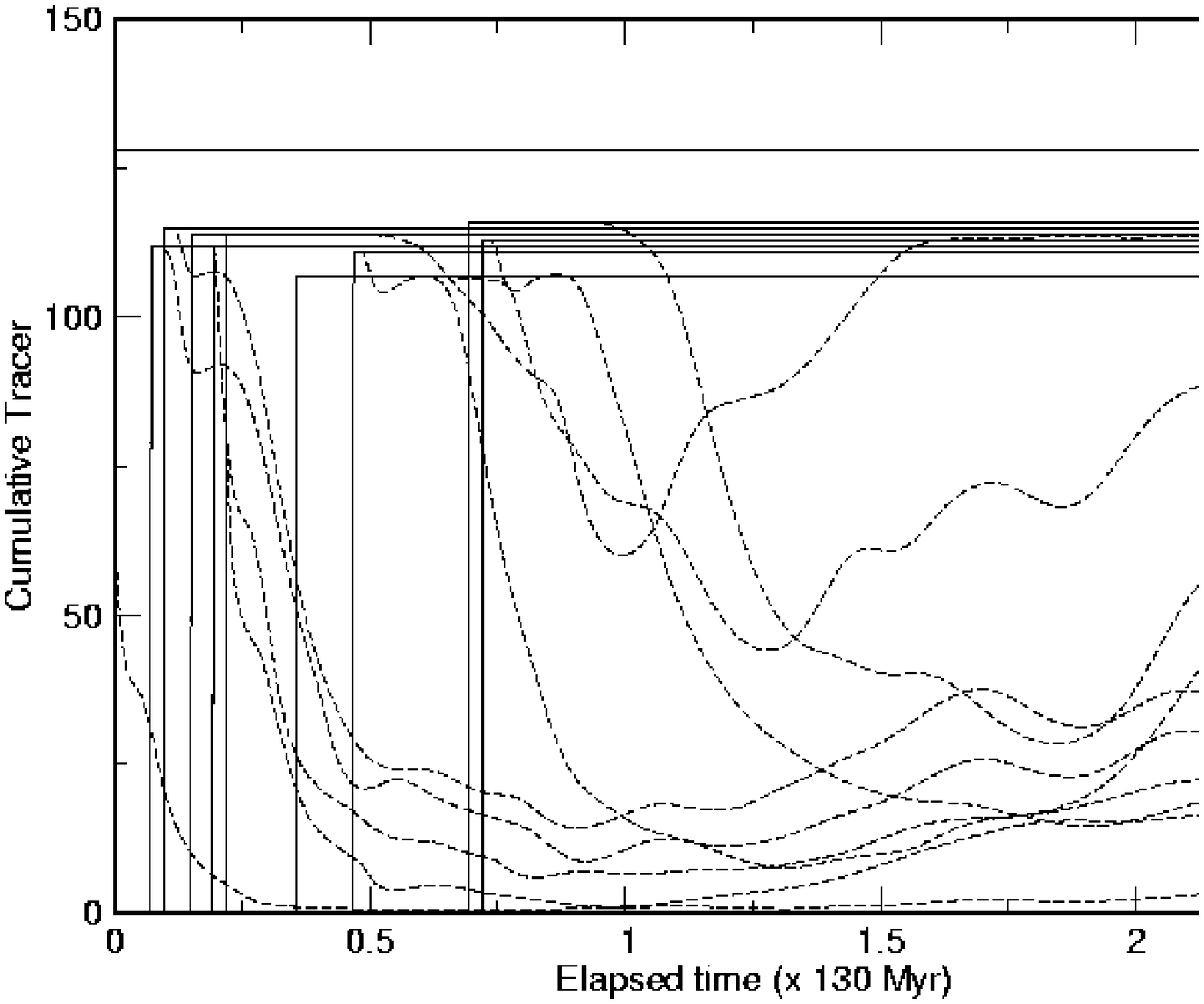}
\caption{
As Figure~\ref{fig:trrad10SNLeo}, but for
(a) Model~E and (b) Model~F.  The results
are displayed as in Figure~\ref{fig:trrad10SNLeo}.
}
\label{fig:trrad10SNFor}
\end{figure}

Models~E and F were similar to C and D, except that the higher mass
galaxy model (Model~2) was used.  The evolution of the tracer fluids 
for these models
is shown in Figure~\ref{fig:trrad10SNFor}, displayed as in
Figure~\ref{fig:trrad10SNLeo}.  The time unit for galaxy Model 2 is 130 Myr.  
A large fraction of the tracer is lost from the core, though less than seen
in Models~C1 and D, and almost no tracer is lost beyond $R_t$.  
Such differences are expected given the
larger mass of this galaxy model.
Little mass loss is seen from the system in Models E and F,
as expected from the tracer evolution.

\subsubsection{The Effects of Numerical Techniques}
\label{sec:numericaleffects}

As discussed above, we tested the effects of resolution upon our results by 
reducing the spatial resolution of Model~C2 by a factor of two relative to  
Model~C1.  In order to
keep other factors as similar as possible, the physical radii over which the
supernova energies were added was kept the same, requiring that the energy
be added over many fewer zones.  The results were similar to Model~C1, 
with
almost identical amounts of tracer loss from the core and overall mass loss.
Somewhat less tracer was lost, however, beyond $R_t$ in the lower resolution
model (33\% as compared to 47\%).  Such a difference would be expected if
the lower resolution allowed more cooling in the supernova-heated gas, reducing
its buoyancy.

In Model C3, cooling was allowed at shock fronts, but was still prevented
during the initial expansion of the hot gas.  As compared to Model~C1,
Model~C3 lost noticeably less tracer and mass beyond $R_t$, with at least
25\% reductions relative to Model~C1.  These results
confirm our suspicion that allowing cooling in the under-resolved shock fronts
results in a noticeable increase in radiative losses, though the effect upon
the evolution is not overly large.

The results of Model~C4, in contrast, suggest that the technique of turning
off the cooling in the initial supernova region until it doubles in radius has
only a small effect upon the final results.  This result implies that the gas
radiates away relatively little energy before the Sedov-Taylor solution is
reached.

In Model~C5, radiative cooling is allowed everywhere and at all times.
As would be expected from the above results, the
results of Model~C5 are very similar to those of Model~C3.

Model~C6 is similar to Model~C1, except that we assume a metallicity for the
gas of $Z=0.3Z_\odot$, ten times larger than that of Model~C1.  As can be seen
in the table, this has no noticeable impact upon the results.  From this, we
conclude that, by the time the post-shock gas temperature drops to below
$10^4$~K, where metal cooling dominates, the evolution of the gas is 
determined by the momentum given to it by the shock, and is no longer 
affected by shock-heating.

In Model~C7, we examine the effects of turning off radiative cooling
completely, i.e. the gas evolves adiabatically from its initial conditions.
As would be expected from earlier work \citep{Larson74}, the differences with
all of the other models are dramatic.  In Model~C7,
most of the original gas in the galaxy ($\approx80$\%) is expelled due to the
energy input of the 10 supernovae, consistent with the estimate of the binding 
energy of galaxy Model~1.  Similarly, almost all of the tracer material is
lost.

From these results, we conclude, consistent with earlier studies, that
radiative losses are crucial to the ability of dwarf galaxies to retain
gas following supernova explosions.  Numerical techniques employed to
improve the accuracy of simulations, by turning off cooling in regions where
overcooling is likely to occur, have quantitative effects upon the results,
but do not affect qualitative conclusions that significant amounts of
both tracer and gas are lost to the systems.  The metallicity of the gas is
found not to be a significant factor.

\subsection{Multiple Supernovae in Systems with an Irregular ISM}
\label{sec:10SNbub}

\begin{figure}
\plotone{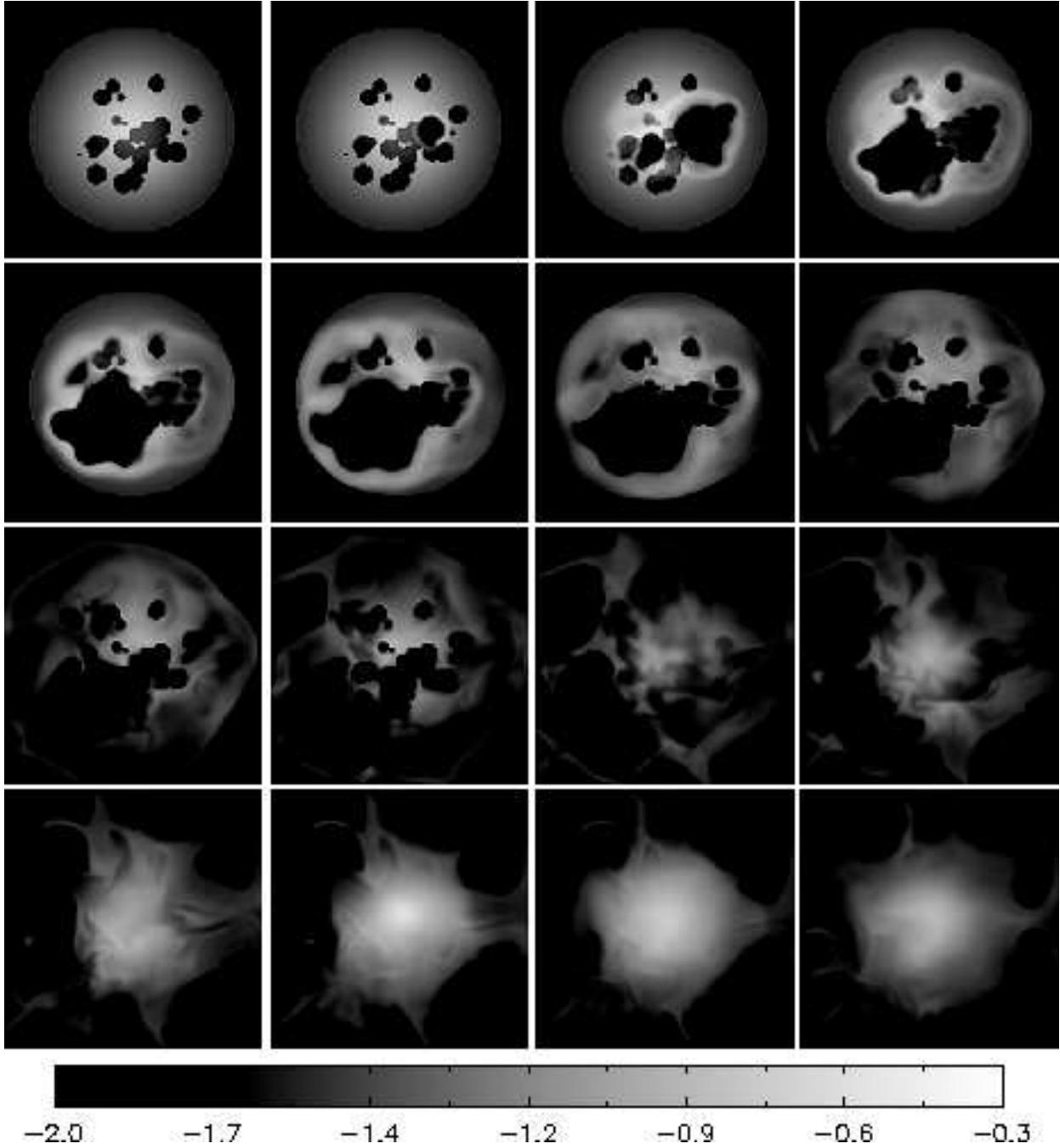}
\caption{Slices through the central plane of the galaxy showing the 
evolution of the logarithm of gas density for Model~G.  The frames correspond
to those shown in Figure \ref{fig:rho10SNLeo10Myr}.}
\label{fig:rhoBubLeo10Myr}
\end{figure}

\begin{figure}
\plotone{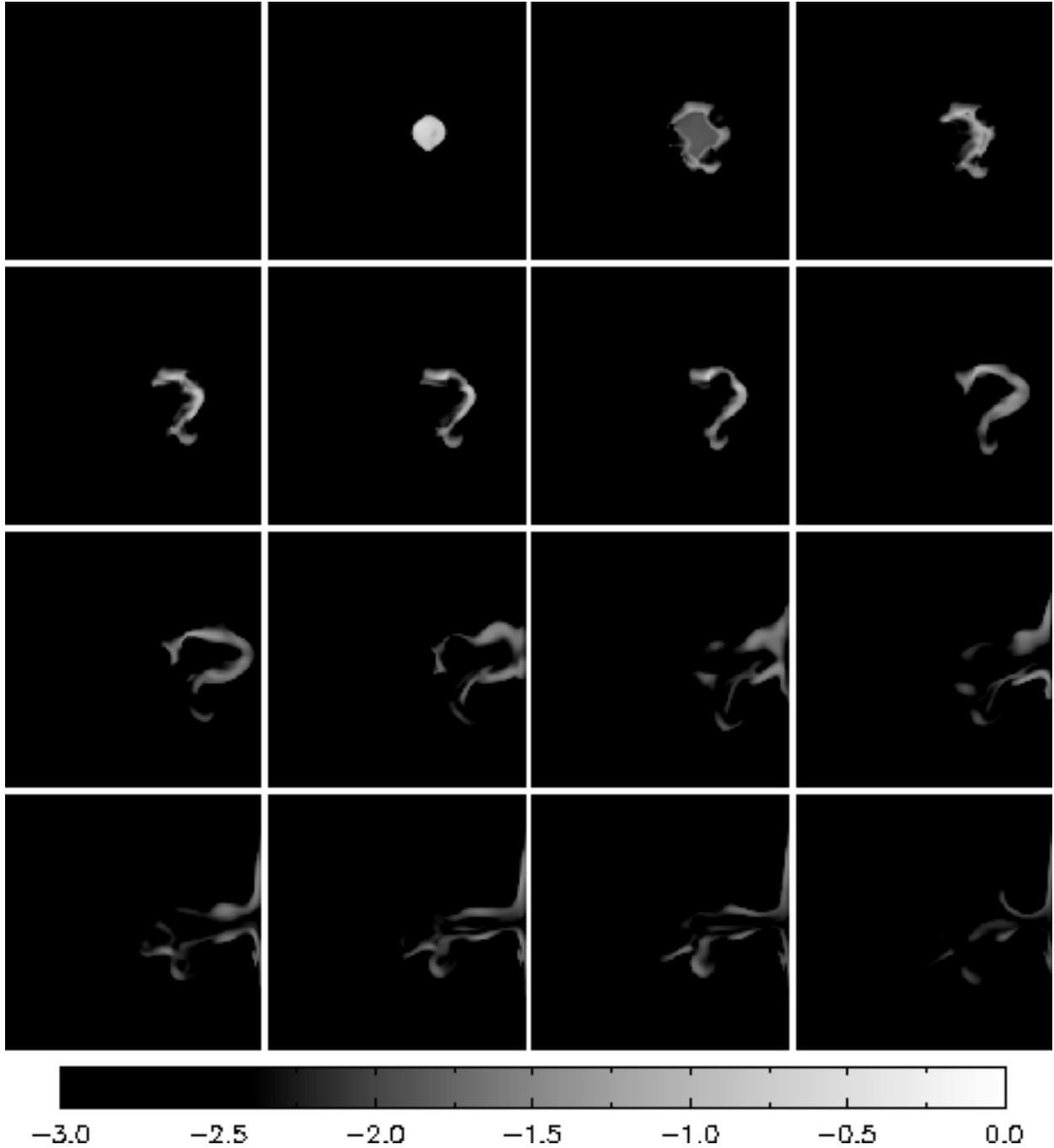}
\caption{Slices through the central plane of the galaxy showing the 
evolution of a single tracer fluid for Model~G.  The frames correspond
to those shown in Figure \ref{fig:rho10SNLeo10Myr}.  Again the tracer fluid
belongs to the first supernova, visible in the second frame of 
Figure \ref{fig:rhoBubLeo10Myr}.}
\label{fig:trBubLeo10Myr}
\end{figure}

Models~G and H use, respectively, the low- and high- mass galaxy models.  As in
Models~C and E, 10 supernovae are triggered over a timescale of 10~Myr.  They
differ from those earlier models in containing 100 low-density 
bubbles, which are intended 
to simulate an irregular ISM, as described in
\S~\ref{sec:methodism}.  In Figure~\ref{fig:rhoBubLeo10Myr} are shown 
snapshots of density evolution in the central galactic plane for Model G, 
covering the same time span as those in Figure~\ref{fig:rho10SNLeo10Myr}.
The frothy nature of the initial gas distribution is clearly illustrated 
in the first few frames.  As described in \S~\ref{sec:method}, the bubbles
are allowed to cool after $\tau_{SN} + 0.5\tau_d$ (34.1 Myr for galaxy Model
1 and 75 Myr for galaxy Model 2).  The cooling evolution is apparent in the 
eleventh frame of Figure~\ref{fig:rhoBubLeo10Myr}.  
In Figure~\ref{fig:trBubLeo10Myr} is shown the evolution of the same tracer
fluid as in Figure~\ref{fig:tr10SNLeo10Myr}, now for Model~G.

\begin{figure}
\plottwo{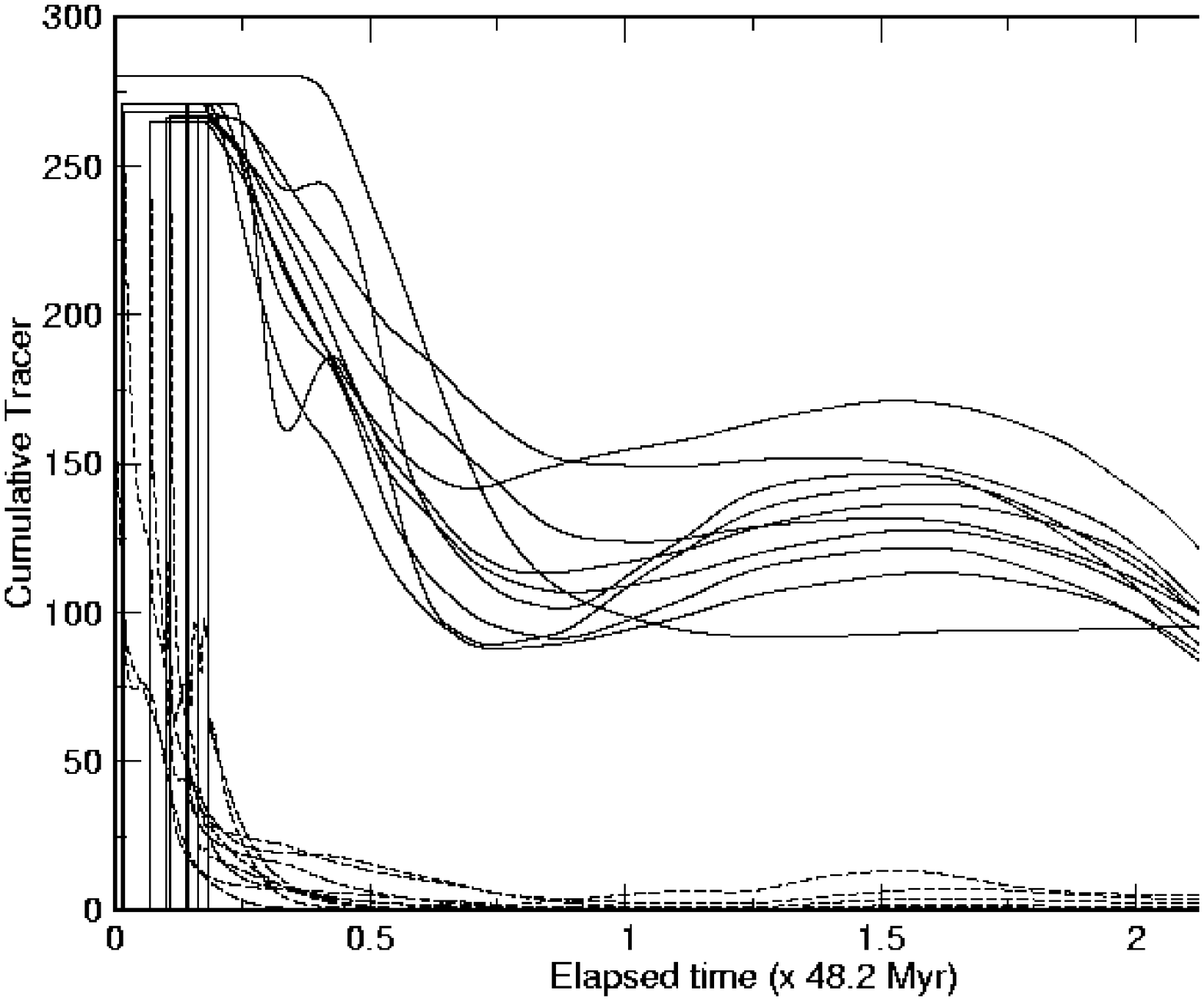}{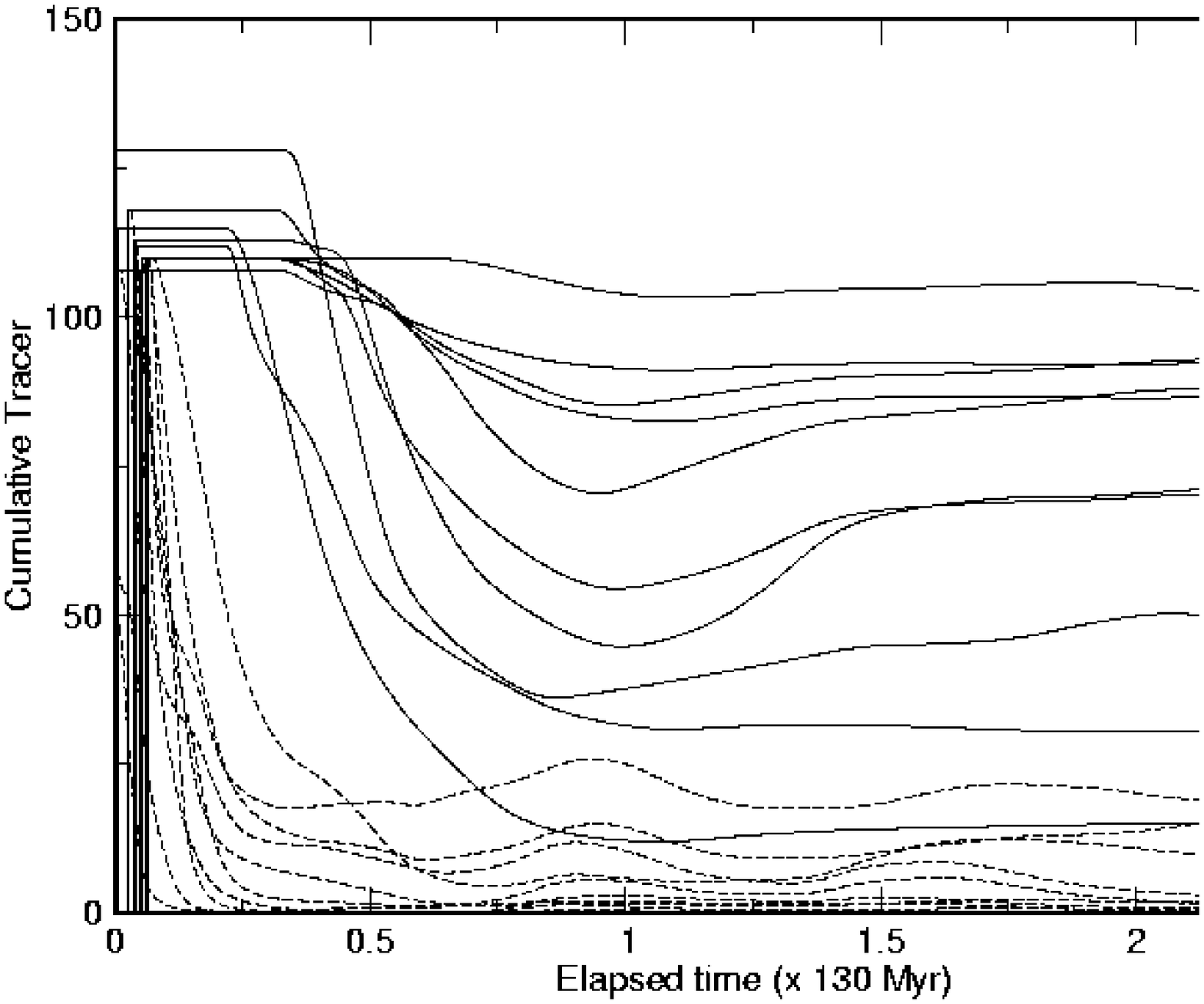}
\caption{
As Figure~\ref{fig:trrad10SNLeo}, but
for Models~G and H.
}
\label{fig:trrad10SNbub}
\end{figure}

The evolution of the tracer and the mass for the two
models are shown in Figures~\ref{fig:trrad10SNbub} and \ref{fig:mass10SNbub}.
The reduction of the average density in the cores of the models, and the
presence of low density channels lead to dramatic differences from the
earlier models.  Nearly all of the enriched material is lost from the cores
of both models.  Approximately 70\% is lost beyond 
$R_t$ as well for Model G, while approximately 50\% is lost beyond $R_t$ 
for Model H.
The systems also show mass losses of 40-50\%.

As in Model~C, the loss of tracer from the cores of Models~G and H is 
accompanied by mass loss.  Unlike the tracer loss, the mass loss later
reverses, again implying
poor mixing of the enriched gas within the galactic cores.  Unlike the
earlier model, the core masses eventually overshoot their initial values.
The overshoot is due to the cooling of the bubbles, which causes some
inflow into the central regions.  As before, the enriched material which is
lost beyond $R_t$ is poorly mixed, as can be seen in 
Figure~\ref{fig:trBubLeo10Myr}.

\section{Differences Between the Models}
\label{sec:differences}

Models which included only a single supernova generally showed little loss
of tracer material.  In the case of supernovae centered in the galaxies, no
tracer was lost even from the core of the galaxy.  In the case of off-center
supernovae, only the low mass system lost some of the enriched gas beyond
the tidal radius.  Such a lack of any effect may be surprising, given the weak
gravitational binding energy of the gas, $\Phi$, in the two model galaxies,
as shown in Table~\ref{tab:galaxies}.  In Model~1, for example,
$\Phi=9.8\times10^{50}$~ergs, comparable to the energy
of a single supernova.  As discussed below, however, the relatively high
gas densities of the systems allow them to radiate away the deposited energy
relatively efficiently, with the result that a single supernova has little
overall effect upon the systems.

The occurrence of multiple supernovae enhances the loss of enriched material in
a few different ways.  
Early supernovae disturb the gas in the core of the galaxies, reducing the
average density.  The cooling efficiency of the gas is therefore reduced, 
as is the mass of gas available to physically contain the ejecta from later
events.  Later events may also help to give a ``boost'' to the enriched material
from earlier events, as can be seen in Figure~\ref{fig:tr10SNLeo10Myr}, 
being especially apparent at the times 3.8 and 7.7~Myr.  Note that the
apparent acceleration in the evolution of the tracer between 24.1 and 33.7~Myr
is an illusion caused by the increasing time interval between frames.

The differences among the models with multiple supernovae are due to
their different masses, and their relative abilities to cool within a 
dynamical timescale.  As shown in Table~\ref{tab:galaxies}, Models~1 and 2
have very similar central densities.  Their different core radii
therefore lead directly to different core masses, as discussed in
\S~\ref{sec:1SN}, with Model~2 having a greater amount of mass available to
contain supernova ejecta than Model~1.

In the absence of cooling, the hot, enriched material from the supernovae 
would still eventually leave the galaxies due to their buoyancy.  Radiative
cooling prevents this, and the different size scales for the two galaxy models
lead to very different ratios of cooling to dynamical times. 
The dynamical timescales from Equation~(\ref{eqn:taudyn}) are
$\tau_d=$~47 and 128 Myr for galaxy Models~1 and 2, respectively.  
The cooling timescale is defined as (Field 1965)
\begin{equation}
\tau_c\sim \frac{\frac{3}{2}k_B T}{n_H L(T,Z)},
\label{eqn:taucool}
\end{equation}
where $L=\Lambda/n_H^2$ is the cooling efficiency in erg~cm$^3$~s$^{-1}$,
and $n_H$ is the number density of hydrogen.  
Close to their initial temperatures, and using our assumed metallicities,
$\tau_c\approx$~36 and 39~Myr, for Models~1 and 2 respectively.  The initial
cooling of the supernova-heated gas shall be much more efficient, with a 
correspondingly shorter timescale.  For the gas to cool to its initial
temperature, however, shall require timescales on the order of those 
calculated above.  For Models~1 and 2, we find, respectively,
$\tau_c/\tau_d=$0.77 and 0.30.  The
values for $\tau_d$ would be smaller, and the timescale ratios larger
by a factor of three had we used $R_c$ instead of $R_t$ in
Equation~(\ref{eqn:taudyn}).

The higher mass galaxy model is therefore able to cool much more on a
dynamical timescale than is the lower mass model.  Following the
conversion of the deposited energy into kinetic energy, the high mass
model can thus radiate away much of the energy before the
enriched material can evolve significantly in space, further enhancing
the ability of the more massive system to retain the enriched gas.
This result is consistent with the observed correlation between 
metallicity and total mass in dSph's \citep{Mateo98}.

The above factors work together to explain the differences seen
between the pair of Models~C1 and D, which use the low-mass galaxy model, and
the pair of Models~E and F, which use the high-mass galaxy model.  In the first
pair, almost all of the enriched gas is lost from within $R_c$, and both
lose a significant amount beyond $R_t$.  Both models
also show a substantial amount of mass loss.  In Models~E and F,
however, less enriched material is lost from within $R_c$, almost none
is lost from $R_t$, and there is much less mass loss. 

The ability of the galaxy to cool between supernova events can also
explain the differences between Models~C1 and D, seen in
Figure~\ref{fig:trrad10SNLeo}.  In Model~C1, the supernovae occur over
a total time of 10~Myr, and the galaxy is unable to cool significantly
between events, whereas the longer 100~Myr supernova timescale of
Model~D allows the system to radiate a significant amount of energy
between supernovae.  As compared to Model~D, then, the supernovae have
a much stronger cumulative effect in Model~C1, leading to more rapid
loss of enriched material from the core, as well as loss beyond $R_t$.
The implication of these models is that star-burst events do not
necessarily lead to greater enrichment of residual gas than does continuous
star formation.

The differences between Models~E and F are less striking than those
between C1 and D, due primarily to the fact that the more massive system is
able to retain enriched material as a whole more efficiently than the less
massive system.  As can be seen in Figure~\ref{fig:trrad10SNFor},
there is less loss from the core of enriched material from a few of the
supernovae in Model~F as compared to Model~E, which primarily accounts for the
quantitative differences between the two models.  Those supernovae happened
to occur at late times, as the system was ``recovering'' from the earlier
supernova events.

Models~G and H were identical, respectively, to Models~C1 and E, except
for the presence of bubbles.  The large filling factors of the bubbles
in the models dramatically altered the evolution of the tracer.  The
bubbles reduce the amount of mass in the core of the galaxy, and
increase the average cooling time of the galaxy.  They also provide
many channels for enriched gas to exit the core.  As a result, most of
the tracer is lost beyond $R_c$ in both models, and a significant
fraction is lost beyond $R_t$.  Both models also lose 40-50\% of the
ambient gas.  This result can be used to infer an upper limit on the
rate of massive star production in order for there to be sufficient
retained gas to fuel subsequent generations of star formation.

\section{Consequences for the Evolution of Dwarf Spheroidal Galaxies}
\label{sec:consequences}

We have performed three-dimensional, numerical simulations of the evolution
of enriched material from supernovae in dwarf spheroidal galaxies.
The simulations have included either one or ten supernovae, and in the case of
multiple supernovae, they have occurred over timescales of either 10 or 100~Myr.
Two different galaxy models, 1 and 2, have been considered, chosen to
approximate the parameters for the Leo~II and Fornax dSph's, respectively.  
Within these galaxy models, both smooth and irregular ISM distributions 
have been considered.  The results of the models, discussed in
\S~\ref{sec:results} and \ref{sec:differences}, are summarized in
Table~\ref{tab:results}.  We now discuss the possible consequences of
our results for the evolution of dSph's.  

Our results indicate that, for extremely low star formation efficiencies, 
such as would lead to individual supernovae widely separated in time, it is
difficult to expel enriched material completely from dwarf spheroidal
galaxies (dSph's) having an undisturbed ISM.  In lower-mass systems,
however, most enriched gas is lost from the cores of the galaxies
following multiple supernovae, and a significant fraction is lost to
the galaxy as a whole.  In a system that has undergone a
recent burst of star formation, however, the ISM is likely to be
highly disturbed.  We find that the presence of an irregular ISM
greatly enhances the loss of enriched material both from the core as
well as from the galaxy as a whole, in both low and high mass models.
In all cases, the material is lost quickly, leaving the core within
$0.2\tau_d$ after the supernovae.  Enriched material that is lost to
the galaxy is typically lost within $2\tau_d$.  As can be seen in the
tracer plots, enriched gas does not mix well prior to being lost to
the galaxy, leading to highly enriched ``blobs'' of material escaping
beyond $R_t$.

The preferential loss of enriched gas from the cores of dSph's might
potentially lead to an ``inverse'' metallicity gradient, the
consequences of which for subsequent star formation depend upon
where later generations of stars form.  If subsequent star formation
is triggered by relatively quiescent processes, such as tidal
compression, or runaway cooling resulting from self-shielding from
external radiation, then star formation is expected to occur primarily
in regions where the gas density is highest, i.e. in the core.  In
such cases, star formation shall occur in gas which is not heavily
enriched from the previous generation of stars, with the result that
multiple generations of stars could be present in the galaxy without
strong metallicity variations.  Given that the ISM is likely to be
highly disturbed by the presence of hot stars, our models indicate
that, in this case, poor self-enrichment efficiency is likely to be
the norm in dSph's.

Star formation triggered by localized compressions may, in contrast,
occur at larger radii, where the gas may have been previously enriched.  
This might
occur, for example, due to collisions between dwarf systems,
encounters with a jet from a nearby massive galaxy, or due to the
infall of material onto the dwarf galaxy.  The latter case might well
be the result of infall of material that was originally expelled from
the dwarf itself.  When material in our simulations is lost from the
galaxies, it exits at below the sound speed
($\lesssim10$~km~s$^{-1}$), much less than the orbital speed of dwarf
galaxies around systems such as the Milky Way.  The ejected gas shall,
therefore, follow a similar orbit around the parent galaxy, and may
collide with gas remaining within the dwarf system due to orbit
crossing at apogalacticon (Dong, Lin, \& Oh 2002).  Such an encounter
might lead to star formation at large radii in the dwarf, and a
subsequent generation of stars showing significant metallicity
enhancement.

Enriched gas which accumulates in the outer regions of dwarf galaxies
may eventually be lost to the galaxy through ram-pressure stripping,
tidal effects, or due to subsequent generations of star formation in
the core.  Thus, as much as 90\% of the supernova ejecta could be
lost, and may enrich the halo of a nearby massive galaxy.  The halo of
the Milky Way, for example, could have been enriched to its average
metallicity by a population of dwarf spheroidals ten times larger than
the number currently observed.

\begin{acknowledgements}
This work was performed under the auspices of the U.S. Department of
Energy by University of California, Lawrence Livermore National
Laboratory under Contract W-7405-Eng-48.  This work is partially
supported by NASA through an astrophysical theory grant NAG5-12151.
\end{acknowledgements}

\clearpage
\begin{deluxetable}{cccccccc}
\tablewidth{0pt}
\tablecaption{Galaxy Parameters \label{tab:galaxies}}
\tablehead{
\colhead{Galaxy} &
\colhead{$M_d$} &
\colhead{$M_g$} &
\colhead{$R_c$} &
\colhead{$\rho_0$} &
\colhead{$\tau_d$} &
\colhead{$\Phi$} &
\colhead{E$_{th}$} \\
\colhead{} &
\colhead{(10$^6$~M$_\odot$)} &
\colhead{(10$^6$~M$_\odot$)} &
\colhead{(pc)} &
\colhead{($10^{-24}$~g~cm$^{-3}$)} &
\colhead{(Myr)} &
\colhead{($10^{51}$~erg)} &
\colhead{($10^{51}$~erg)}
}
\startdata
1  & $10$ & $1.0$ & 130 & $1.2$ & 47 & 0.98 & 2.5 \\
2  & $55$ & $5.5$ & 360 & $1.1$ & 128 & 16. & 12.
\enddata
\end{deluxetable}

\clearpage
\begin{deluxetable}{ccccc}
\tablewidth{0pt}
\tablecaption{Model Parameters \label{tab:models}}
\tablehead{
\colhead{Model} &
\colhead{Galaxy} &
\colhead{$N_{SN}$} &
\colhead{$\tau_{SN}$} &
\colhead{$N_{\rm{bub}}$} \\
\colhead{} &
\colhead{} &
\colhead{} &
\colhead{(Myr)} &
\colhead{}
}
\startdata
A  & 1 & 1 & -- & 0 \\
B  & 2 & 1 & -- & 0 \\
C  & 1 & 10 & 10 & 0 \\
D  & 1 & 10 & 100 & 0 \\
E  & 2 & 10 & 10 & 0 \\
F  & 2 & 10 & 100 & 0 \\
G  & 1 & 10 & 10 & 100 \\
H  & 2 & 10 & 10 & 100 
\enddata
\end{deluxetable}

\clearpage
\begin{deluxetable}{cccc}
\tablewidth{0pt}
\tablecaption{Results of the Models\label{tab:results}}
\tablehead{
\colhead{Model} &
\colhead{Core tracer loss} &
\colhead{Total tracer loss} &
\colhead{Mass loss} \\
\colhead{} &
\colhead{\%} &
\colhead{\%} &
\colhead{\%} 
}
\startdata
A1 & 0 & 0 & 10 \\
A2 & 92 & 30 & 7 \\
B1 & 0 & 0 & 3 \\
B2 & 98 & 0 & 2 \\
C1 & 94 & 47 & 36 \\
C2 & 93 & 33 & 33 \\
C3 & 90 & 30 & 27 \\
C4 & 94 & 50 & 35 \\
C5 & 91 & 29 & 26 \\
C6 & 94 & 47 & 36 \\
C7 & 100 & 84 & 80 \\
D & 79 & 19 & 21 \\
E & 65 & 3 & 10 \\
F & 59 & 0 & 7 \\
G & 99 & 71 & 51 \\
H & 94 & 46 & 37 
\enddata
\end{deluxetable}


\clearpage

\clearpage

\clearpage

\clearpage

\clearpage

\clearpage

\clearpage

\clearpage

\clearpage

\clearpage
\begin{figure}
\plottwo{f10a.eps}{f10b.eps}
\caption{
As Figure~\ref{fig:mass10SNLeo}, but
for Models~G and H. }
\label{fig:mass10SNbub}
\end{figure}


\clearpage
\begin{thebibliography}{}

\bibitem[Andersen \& Burkert (2000)]{AB00} Andersen, R.-P., \& Burkert,
A. 2000, \apj, 531, 296

\bibitem[Anninos \& Fragile (2003)]{AF03} Anninos, P., \& Fragile, P. C. 2003, 
\apjs, 144, 243

\bibitem[Anninos, Fragile \& Murray (2002)]{AFM02} Anninos, P., Fragile,
P. C., \& Murray, S. D. 2002, \apj, submitted

\bibitem[Anninos \& Norman (1994)]{AN94} Anninos, W. Y., \& Norman, M. L. 1994,
\apj, 429, 434

\bibitem[Babul \& Rees (1992)]{BR92} Babul, A. \& Rees, M. J. 1992, 
\mnras, 255, 346

\bibitem[Barkana \& Loeb (1999)]{BL99} Barkana, R., \& Loeb, A. 1999,
\apj, 523, 54

\bibitem[Benson et al. (2002)]{BLBCF02} Benson, A. J., Lacey, C. E.,
Baugh, C. M., Cole, S., \& Frenk, C. S. 2002, \mnras, 333, 156

\bibitem[Binney \& Tremaine (1987)]{BT87} Binney, J. \& Tremaine, S. 1987,
Galactic Dynamics (Princeton: Princeton Univ. Press)

\bibitem[Blondin, Fryxell, \& Konigl (1990)]{BFK90} Blondin, J. M., Fryxell,
B. A., \& Konigl, A. 1990, \apj, 360, 370

\bibitem[Blumenthal et al. (1984)]{BFPR84} Blumenthal, G. R., Faber, S. M.,
Primack, J. R., \& Rees, M. J. 1984, \nat, 311, 517

\bibitem[Burkert (2000)]{B00} Burkert, A. 2000, \apj, 534, L143

\bibitem[Burkert (1995)]{B95} Burkert, A. 1995, \apj, 447, L25

\bibitem[Cole et al. (1994)]{C94} Cole, S., Aragon-Salamanca, A., Frenk, C. S.,
Navarro, J. F., \& Zepf, S. E. 1994, \mnras, 271, 781

\bibitem[C\^ot\'e, Oke, \& Cohen (2000)]{Cote00} C\^ot\'e, P., Oke, J. B., 
\& Cohen, J. G. 2000, \aj, 118, 1645

\bibitem[Dekel \& Silk (1986)]{DS86} Dekel, A., \& Silk, J. 1986,
\apj, 303, 39

\bibitem[Dohm-Palmer et al. (1998a)]{DPetal98a} Dohm-Palmer, R. C., Skillman,
E. D., Gallagher, J. S., Tolstoy, E., Mateo, M., Dufour, R. J., Saha, A., 
Hoessel, J., \& Chiosi, C. 1998, \aj, 116, 1227

\bibitem[Dohm-Palmer et al. (2002)]{DPetal02} Dohm-Palmer, R. C., Skillman,
E. D., Mateo, M., Saha, A., Dolphin, A., Tolstoy, E., Gallagher, J. S.
\& Cole, A. A. 2002, \aj, 123, 813

\bibitem[Dohm-Palmer et al. (1998b)]{DPetal98b} Dohm-Palmer, R. C., Skillman,
E. D., Saha, A., Tolstoy, E., Mateo, M., Gallagher, J. S., Hoessel, J., 
Chiosi, C., \& Dufour, R. J. 1998, \aj, 114, 2527

\bibitem[Dong, Lin, \& Oh (2002)]{DLO02} Dong, S., Lin, D. N. C., \&
Oh, K. S. 2002, in preparation

\bibitem[Dong, Murray, \& Lin (2003)]{DML03} Dong, S., Murray, S. D.,
\& Lin, D., N. C. 2002, \apj, submitted

\bibitem[Efstathiou (1992)]{E92} Efstathiou, G. 1992, \mnras, 256, 43P

\bibitem[Field (1965)]{F65} Field, G. B. 1965, \apj, 142, 531

\bibitem[Gallagher et al. (1998)]{Getal98} Gallagher, J. S., Tolstoy, E.,
Dohm-Palmer, R. C., Skillman, E. D., Cole, A. A., Hoessel, J. G., Saha,
A., \& Mateo, M. 1998, \aj, 115, 1869

\bibitem[Grebel (2001)]{Grebel01} Grebel, E. K. 2001, ASSS, 277, 231

\bibitem[Grebel (1997)]{Grebel97} Grebel, E. K. 1997, Rev. Mod. Astron.,
10, 29

\bibitem[Harbeck et al. (2001)]{H01} Harbeck, D., Grebel, E. K., Holtzmann, J.,
Guhathakurta, P., Brandner, W., Geisler, D., Sarajedini, A., Dolphin, A.,
Hurley-Keller, D., \& Mateo, M. 2001, \aj, 122, 3092

\bibitem[Heiles (1987)]{Heiles87} Heiles, C. 1987, Interstellar Processes,
ed. D. J. Hollenbach \& H. A. Thronson, (Dordrecht: D. Reidel), 171

\bibitem[Kepner, Babul, \& Spergel (1997)]{KBS97} Kepner, J. V., Babul, A.,
\& Spergel, D. N. 1997, \apj, 487, 61

\bibitem[Klypin et al. (1999)]{K99} Klypin, A., Kravtsov, A. V., 
Valenzuela, O., \& Prada, F. 1999, \apj, 522, 82

\bibitem[Klypin, Nolthenius, \& Primack (1997)]{KNP97} Klypin, A., Nolthenius,
R., \& Primack, J. 1997, \apj, 474, 533

\bibitem[Larson (1974)]{Larson74} Larson, R. B. 1974, \mnras, 169, 229

\bibitem[Lehnert et al. (1992)]{Lehnert92} Lehnert, M. D., Bell, R. A.,
Hesser, J. E., \& Oke, J. B. 1992, \apj, 395, 466

\bibitem[Lin \& Murray (1994)]{LM94} Lin, D. N. C., \& Murray, S. D.
1994, Dwarf Galaxies (Garching: ESO), 535

\bibitem[Mac Low \& Ferrara (1999)]{MLF99} Mac Low, M-M., \& Ferrara, A. 1999,
\apj, 513, 142

\bibitem[Mac Low \& McCray (1988)]{MLM88} Mac Low, M-M., \& McCray, R. 1988,
\apj, 324, 776

\bibitem[Madau, Ferrara, \& Rees (2001)]{MFR01} Madau, P., Ferrara, A., \&
Rees, M. J. 2001, \apj, 555, 92

\bibitem[Mateo (1998)]{Mateo98} Mateo, M. 1998, \araa, 36, 435

\bibitem[Mori \& Burkert (2000)]{MB00} Mori, M., \& Burkert, A. 2000,
\apj, 538, 559

\bibitem[Mori, Ferrara, \& Madau (2002)]{MFM02} Mori, M., Ferrara, A., \& 
Madau, P. 2002, \apj, 571, 40

\bibitem[Mori et al. (1997)]{M97} Mori, M., Yoshii, Y., Tsujimoto, T.,
\& Nomoto, K. 1997, \apj, 478, L21

\bibitem[Murray, Dong \& Lin (2003)]{MDL03} Murray, S. D., Dong, S., \& Lin,
D. N. C. 2003, \apj, submitted

\bibitem[Navarro, Frenk \& White (1997)]{NFW97} Navarro, J. F., Frenk, C. S.,
\& White, S. D. M. 1997, \apj, 490, 493

\bibitem[Navarro \& Steinmetz (1997)]{NS97} Navarro, J. F., \& Steinmetz,
M. 1997, \apj, 478, 13

\bibitem[Navarro \& White (1994)]{NW94} Navarro, J. F., \& White, S. D. M.
1994, \mnras, 267, 401

\bibitem[Noriega-Crespo et al. (1989)]{NBLT89} Noriega-Crespo, A., 
Bodenheimer, P., Lin, D. N. C., \& Tenorio-Tagle, G. 1989, \mnras, 237, 46

\bibitem[Peebles \& Dicke (1968)]{PD68} Peebles, P. J. E., \& Dicke, R. H.
1968, \apj, 154, 891

\bibitem[Persic, Salucci, \& Stel (1996)]{PSS96} Persic, M., Salucci, P.,
\& Stel, F. 1996, \mnras, 281, 27

\bibitem[Quinn, Katz, \& Efstathiou (1996)]{QKE96} Quinn, T., Katz, N.,
\& Efstathiou, G. 1996, \mnras, 278, L49

\bibitem[Rosner \& Tucker (1989)]{RT89} Rosner, R. \& Tucker, W. H. 1989,
\apj, 338, 761

\bibitem[Shapiro \& Struck-Marcell (1985)]{SSM85} Shapiro, P. R.,
\& Struck-Marcell, C. 1985, \apjs, 57, 205

\bibitem[Shetrone, C\^ot\'e, \& Sargent (2001)]{Shetrone01} Shetrone, M.
D., C\^ot\'e, P., \& Sargent, W. L. W. 2001, \apj, 548, 592

\bibitem[Smecker-Hane et al. (1996)]{S96} Smecker-Hane, T. A., Stetson, P.
B., Hesser, J. E., \& VandenBerg, D. A. 1996, in From Stars to Galaxies:
The Impact of Stellar Physics on Galaxy Evolution, ed. C. Leitherer, U.
Fritze-von Alvensleben, \& J. Huchra (San Francisco: ASP), 328

\bibitem[Sommer-Larsen, Gelato \& Vedel (1999)]{SLGV99} Sommer-Larsen, J.,
Gelato, S., \& Vedel, H 1999, \apj, 519, 501

\bibitem[Spitzer (1956)]{Spitzer56} Spitzer, L. 1956, Physics of Fully Ionized
Gases (New York: Interscience)

\bibitem[Suntzeff et al. (1993)]{Suntzeff93} Suntzeff, N. B., Mateo, M.,
Terndrup, D. M., Olszewski, E. W., Geisler, D., \& Weller, W. 1993, \apj,
418, 208

\bibitem[Tenorio-Tagle et al. (1986)]{TBLN86} Tenorio-Tagle, G.,
Bodenheimer, P., Lin, D. N. C., \& Noriega-Crespo, A. 1986, \mnras,
221, 635

\bibitem[Weil, Eke \& Efstathiou (1998)]{WEE98} Weil, M. L., Eke, V. R., \&
Efstathiou, G. 1998, \mnras, 300, 773

\bibitem[Weinberg, Hernquist \& Katz (1997)]{WHK97} Weinberg, D. H., Hernquist,
L., \& Katz, N. 1997, \apj, 477, 8

\bibitem[White \& Rees (1978)]{WR78} White, S. D. M., \& Rees, M. J. 1978,
\mnras, 183, 341

\end{thebibliography}
\end{document}